# Fiber Bragg Gratings Embedded in 3D-Printed Scaffolds

Peter Liacouras[1], CAPT Gerald Grant[1], Khazar Choudhry[2], G. F. Strouse[2] and Zeeshan Ahmed[2]

[1]*3D Medical Applications Center, Department of Radiology*
*Walter Reed National Military Medical Center*
*8901 Wisconsin Avenue, Bethesda, MD 20889, USA*

[2]*Thermodynamic Metrology Group, Sensor Science Division, Physics Laboratory, National Institute of Standards and Technology, Gaithersburg, MD 20899*

**Abstract:**

In recent years there has been considerable interest in utilizing embedded fiber optic based sensors for fabricating smart materials. One of the primary motivations is to provide real-time information on the structural integrity of the material so as to enable proactive actions that prevent catastrophic failure. In this preliminary study we have examined the impact of embedding on the temperature-dependent response of fiber Bragg gratings (FBG). Our results indicate that the embedding has a significant impact on the sensor's temperature response suggesting quantitative use of embedded sensors will likely require the development of *in corpus* calibration procedures.

Structural monitoring of infrastructure is a time-consuming, expensive and often dangerous endeavor that requires human inspectors to make frequent visual observations under challenging conditions [1-6]. It is estimated that the United States requires nearly a $1 trillion investment in infrastructure upkeep and replacement [7]. Given the size of this challenge, it is not surprising that there has been a surge of interest in the development of intelligent materials, i.e., materials with embedded sensors that can enable either real-time or on-demand structural health and integrity monitoring [5]. In recent years, fiber optic sensors have emerged as a promising technology that can be easily embedded into a large structure, creating a wide-mesh network that can provide structural integrity information during manufacturing and service, effectively replacing the need for frequent visual inspections. An interesting potential application is in cure monitoring of composite materials where fiber optic-based sensors utilizing either transmission spectroscopy, evanescent wave spectroscopy, refractive index monitoring or fiber Bragg grating (FBG) are employed as temperature/strain sensors [5, 6].

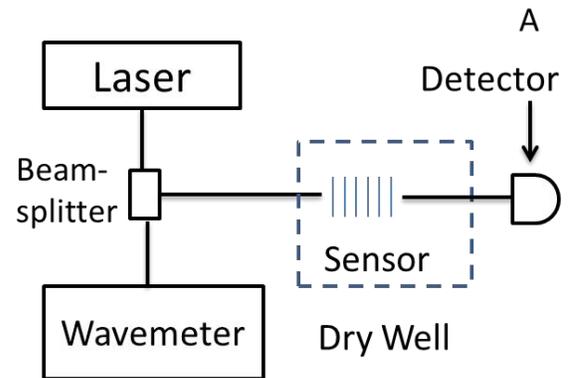

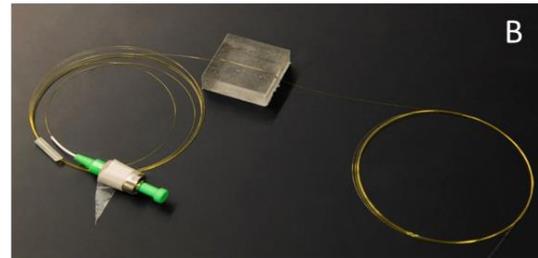

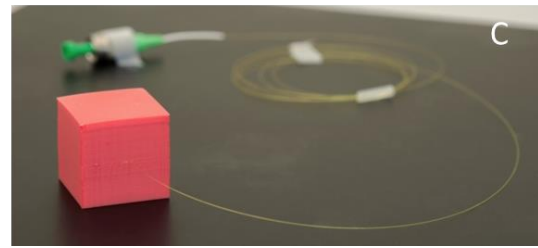

Figure 1: a) Schematic of the experimental setup. FBG sensors embedded into 3D printed (b) Valero and (c) ABS scaffold

FBG is a narrow band optical filter fabricated using photo-sensitive optical fibers (H$_2$ loaded *p*-Ge doped fibers) that are exposed to a spatially varying UV light source such as a deep-UV laser based Michelson interferometer [8]. The periodic varying light induces photo-chemical reactions which modify the local structure of silica to create a periodic variation in the local refractive index that acts like a Bragg grating. The Bragg wavelength is given by:

$$\lambda_B = 2n_e L \quad (1)$$

Where, $\lambda_B$ is Bragg wavelength, $n_e$ is the effective refractive index, and $L$ is the grating period. The wavelength of light resonant with the Bragg period is reflected back, while non-resonant wavelengths pass through the grating [8]. Changes in the surrounding temperature impact the effective grating period by impacting the linear thermal expansion of the material, and/or a change in the fiber's refractive index due to temperature (thermo-optic effect).

$$\lambda(T) = \lambda(R) + \left[2\left(L\left(\frac{\partial n}{\partial T}\right) + n\left(\frac{\partial L}{\partial T}\right)\right)\right](T-R) \quad (2)$$

where, T is the sensor temperature, R is reference temperature and $\lambda(T)$ and $\lambda(R)$ are the Bragg wavelengths at temperatures T and R (reference temperature), respectively. $\frac{\partial L}{\partial T}$ is the linear thermal expansion of silica, while $\frac{\partial n}{\partial T}$ is the temperature variation of refractive index. In fused silica, the thermo-optic effect is a factor of ten larger than thermal expansion. Existing literature indicates the FBG shows a temperature dependent shift of 10 pm/°C, while the strain response is only order of 1 pm/με [8-10].

In this study we have examined the impact on sensor response of embedding sensors into 3D printed scaffolds. We utilize a custom built laser-based FBG interrogation system for recording the FBG response (Fig 1). Briefly, a C-band laser (New Focus TLB-6700 series[11]) is swept over the sensor resonance. A small amount of laser power was immediately picked up from the laser output for wavelength monitoring (HighFinesse WS/7[11]) while the rest, after passing through the photonic device, was detected by a large sensing-area power meter (Newport, model 1936-R[11]). Three FBGs (acquired from Smart Fibers [11] and Micron Optics[11]) were used in this study. Two of the sensors show grating resonances at 1550.087 nm and 1549.73 nm (sensor 1 and 2), while the third sensor shows a resonance at 1552.037 nm. The first two sensors were embedded in a 25 mm x 25 mm x 10 mm block of a photopolymer (Vero Clear Full Cure 810 [11]) fabricated using an Objet Connex 500 printer[11], while the third was embedded in a 1 inch acrylonitrile butadiene styrene (ABS) cube-like scaffold (75% infill) fabricated using a Makerbot Replicator 2X[11] machine. The photopolymer block was created in two separate parts. The first part was a 5 mm thick with a 0.6 mm radius channel across the midline of the top surface. After completion, the final layer number was recorded and the FBG sensor was laid into the channel and taped on both sides to the platform. Finally, the full 10 mm thick block was placed on the platform in the exact same location as the first. This formation was then started one layer later than then the previous build. The ABS scaffold was paused halfway during fabrication to allow FBG sensor to be laid across the scaffold. The FBG was covered with a small piece of tape to protect the fiber from the printer nozzle during printing of the rest of scaffold. As shown in Fig 2, for sensors 1 and 2 being embedded into a 3D printed scaffold results in a large redshift of 29 pm and 4350 pm, respectively. Assuming temperature response of 10 pm/°C, this shift in Bragg wavelength would correspond to temperature change of 2.9 °C and 435 °C, clearly indicating the process of embedding a FBG can introduce large uncertainties in temperature measurements.

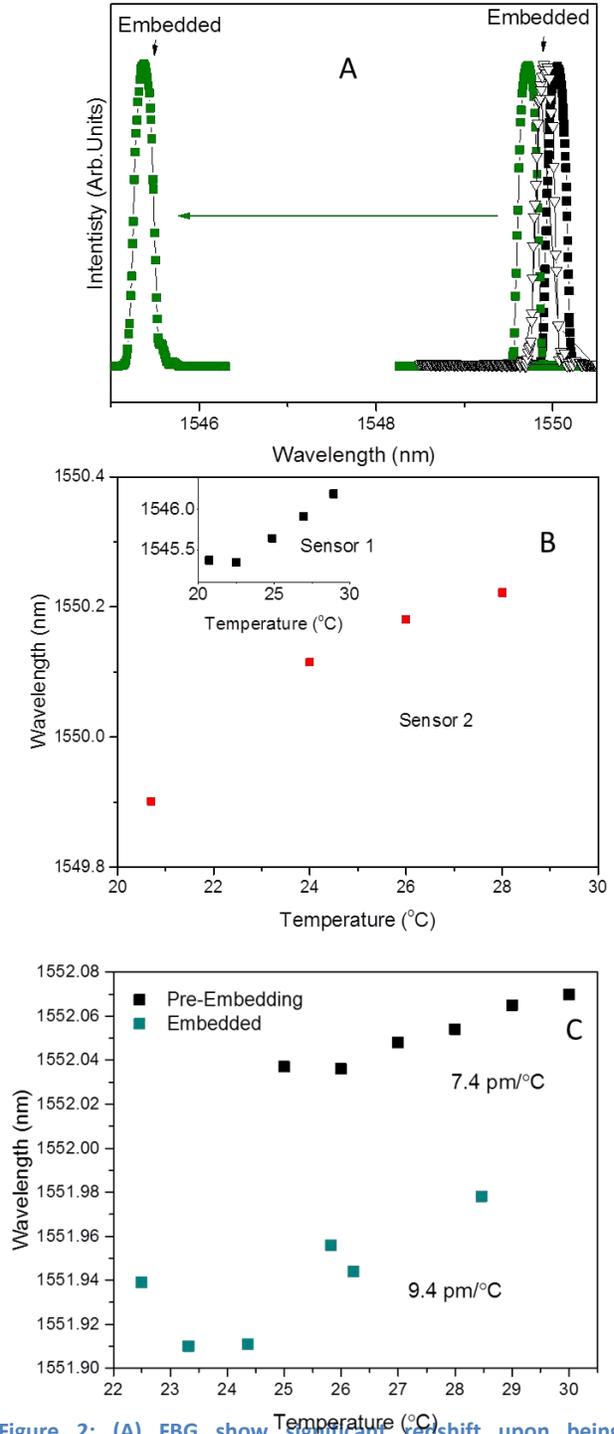

Figure 2: (A) FBG show significant redshift upon being embedded. B) Sensors embedded into Valero scaffold show dramatically different temperature response. C) FBG sensor embedded into ABS scaffold shows a 27% increase in temperature response.

Upon heating, we observe of these embedded sensors, a temperature dependent upshift of 44.5 pm/°C (sensor 1) and 105.3 pm/°C (sensor 2). A more muted behavior is observed for sensor 3, which was embedded in ABS scaffold that was 25% air. Here we observe a temperature-dependent shift of 7.4 pm/°C before embedding and 9.4 pm/°C post-embedding in the ABS scaffold (Fig 2c). The two temperature response curves (before and after embedding) are offset by ≈110 pm or ΔT ≈11 °C. While the zero off-set in calibration curves can be easily recognized and corrected for, the dramatic difference in the wavelength-temperature curve seen in our measurements clearly indicates that any quantitative use of embedded FBG would require *in corpus* calibration.

In summary our examination of FBG embedded in 3D printed scaffolds indicates the surrounding matrix (infill percentage, residual stress, etc.) has a significant impact on sensor's temperature response. Likely due to stress of surrounding matrix, the embedded sensors show significant zero off-sets and variations in temperature-wavelength response that vary from 9.4 pm/°C to 435 pm/°C. Our results suggest that quantitative use of embedded sensors will require the development of *in corpus* calibration procedures.

**Acknowledgement:** Khazar Choudhry was supported by ACS SEED fellowship program.